\documentclass[longauth]{aa} 
\usepackage{graphicx}
\usepackage{graphics}
\usepackage{colortbl}
\usepackage{txfonts}
\usepackage{natbib}

\begin{document}
  \title{Detection of extended very-high-energy $\gamma$-ray emission towards the young stellar cluster Westerlund~2}

\author{F. Aharonian\inst{1}
 \and A.G.~Akhperjanian \inst{2}
 \and A.R.~Bazer-Bachi \inst{3}
 \and M.~Beilicke \inst{4}
 \and W.~Benbow \inst{1}
 \and D.~Berge \inst{1} \thanks{now at CERN, Geneva, Switzerland}
 \and K.~Bernl\"ohr \inst{1,5}
 \and C.~Boisson \inst{6}
 \and O.~Bolz \inst{1}
 \and V.~Borrel \inst{3}
% ...
}
\offprints{Olaf Reimer (olr@stanford.edu), Martin Raue (martin.raue@desy.de)}

\institute{
Max-Planck-Institut f\"ur Kernphysik, P.O. Box 103980, D 69029
Heidelberg, Germany
\and
 Yerevan Physics Institute, 2 Alikhanian Brothers St., 375036 Yerevan,
Armenia
\and
Centre d'Etude Spatiale des Rayonnements, CNRS/UPS, 9 av. du Colonel Roche, BP
4346, F-31029 Toulouse Cedex 4, France
\and
Universit\"at Hamburg, Institut f\"ur Experimentalphysik, Luruper Chaussee
149, D 22761 Hamburg, Germany
\and
Institut f\"ur Physik, Humboldt-Universit\"at zu Berlin, Newtonstr. 15,
D 12489 Berlin, Germany
\and
LUTH, UMR 8102 du CNRS, Observatoire de Paris, Section de Meudon, F-92195 Meudon Cedex,
France
% ...
}


  \abstract {}
      { Results from $\gamma$-ray observations by the H.E.S.S. telescope array in the direction of 
	the young stellar cluster Westerlund~2 are presented.}
      { Stereoscopic imaging of Cherenkov light emission of $\gamma$-ray induced showers 
	in the atmosphere is used to study the celestial region around the massive Wolf-Rayet (WR) 
	binary WR~20a. Spectral and positional analysis is performed using standard event reconstruction 
	techniques and parameter cuts.}
      { The detection of a new $\gamma$-ray source is reported from H.E.S.S. observations 
	in 2006. HESS~J1023--575 is found to be coincident with the young stellar cluster Westerlund~2
	in the well-known HII complex RCW~49. The source is detected with a statistical significance of 
	more than 9$\sigma$, and shows extension beyond a point-like object within the H.E.S.S. point-spread function. 
	The differential $\gamma$-ray spectrum of the emission region is measured over approximately two orders of magnitude in flux.}
      {The spatial coincidence between HESS~J1023--575 and the young open cluster Westerlund~2, 
	hosting e.g. the massive WR binary WR~20a, requires one to look into a variety of potential models 
	to account for the observed very-high-energy (VHE) $\gamma$-ray emission. Considered emission scenarios 
	include emission from the colliding wind zone of WR~20a, collective stellar winds from the 
	extraordinary ensemble of hot and massive stars in the stellar cluster Westerlund~2, diffusive shock 
	acceleration in the wind-blown bubble itself, and supersonic winds breaking out into the interstellar medium (ISM).
	The observed source extension argues against a single star origin of the observed VHE emission.  
	}
	
  \authorrunning{F. Aharonian et al.}  
  \titlerunning{VHE $\gamma$-ray emission from Westerlund~2}

  \maketitle

\section{Introduction}

The prominent giant HII region RCW~49 (also NGC~3247), and its ionizing cluster  
Westerlund~2 are located towards the outer edge of the Carina arm of our Milky Way. 
RCW~49 is a luminous, massive star formation region, and has been extensively 
studied at various wavelengths. After the initial report based on photographic 
images by~\citet{Rod60}, the region was observed at X-ray energies with 
$\it Einstein$~\citep{Her84,Gol87}, ROSAT~\citep{Bel94}, and recently $\it Chandra$, 
which discovered $\sim$500 point sources in the vicinity of RCW~49~\citep{Tsu04}, 
with $\sim$100 of them spatially coincident with the central open stellar 
cluster Westerlund~2~\citep{Tow05}. Mid-infrared measurements with $\it Spitzer$ 
revealed still ongoing massive star formation in RCW~49~\citep{Whi04}. The 
regions surrounding Westerlund~2 appear evacuated by stellar winds and radiation. 
The surrounding dust is distributed in fine filaments, knots, pillars, bubbles, and 
bow shocks throughout the rest of the HII complex~\citep{Chu04, Con04}. Radio continuum 
observations by ATCA at 1.38 and 2.38 GHz indicate two wind-blown shells in the core of 
RCW~49~\citep{Whi97}: one surrounding the central region of Westerlund~2, the other the 
prominent WR star WR~20b. The distance to RCW~49 is still uncertain and values range 
between $\sim$2.2 kpc~\citep{Bra93} up to 7.9 kpc~\citep{Mof91}, whereas intermediate 
values of 4.2 kpc from 21 cm absorption line profile measurements~\citep{McC01}, 
5.75 kpc from the distance estimate towards the prominent WR star WR~20a~\citep{vdH01}, 
and 6.4 kpc from photometric measurements~\citep{Car04} correspond to more recent refinements.
Finally, \citet{Rau07} presented a compelling re-determination of the distance to Westerlund~2 
by spectro-photometric measurements of 12 cluster member O-type stars of (8.3$\pm$ 1.6)~kpc.
This value is in very good agreement with the (8.0$\pm$1.0)~kpc as measured by \citet{Rau05}, 
determined from the light curve of the eclipsing binary WR~20a. In summary, we adopt the distance 
value of the weighted mean of (8.0$\pm$1.4)~kpc \citep{Rau07} throughout the manuscript,
thereby associating WR~20a as a cluster member of Westerlund~2.
  
Westerlund~2~\citep{Wes60} contains an extraordinary ensemble of hot and massive stars, 
presumably at least a dozen early-type O-stars, and two remarkable WR stars. 
Two Wolf-Rayet (WR) stars have been found within RCW~49, WR~20b at $\alpha_{2000}$ = $10^{\rm h}24^{\rm m}18.^{\rm s}4$, 
$\delta_{2000}$ = -57$^\circ$48'30'', and WR~20a at $\alpha_{2000}$ = $10^{\rm h}23^{\rm m}58.^{\rm s}0$, 
$\delta_{2000}$ = -57$^\circ$45'49''~\citep{vdH01}, both first reported as WR stars by \citet{Sha91}, 
the latter believed to be a member of the stellar cluster Westerlund~2. Only recently was WR20a established 
to be a binary: both \citet{Rau04b} and \citet{Bon04} presented solutions for a circular orbit 
with a period of 3.675, and 3.686 days, respectively. Based on the the orbital period, the minimum masses 
have been found to be $(83 \pm 5)$\,M$_{\odot}$ and $(82 \pm 5)$\,M$_{\odot}$ for the primary and secondary 
components, respectively~\citep{Rau05}. This classifies the WR binary WR~20a as the most massive of all confidently 
measured binary systems in our Galaxy. An orbital inclination of $\sim$75$^\circ$ was inferred 
from photospheric eclipses~\citep{Bon04}, and the spectral classification was subsequently 
refined as a binary consisting of two WN6ha stars~\citep{Rau05}. The supersonic stellar winds 
of both WR stars collide, and a wind-wind interaction zone forms at the stagnation point with a 
reverse and forward shock. In a detached binary system like WR~20a, the colliding wind zone lies 
between the two stars, and is heavily skewed by Coriolis forces. The wind velocity at the 
stagnation point is likely lower than in other positions due to the \emph{radiative inhibition mechanism}~\citep{Ste94} 
caused by the companion's stellar photon field. In a close binary system the winds can only be 
radiatively accelerated to a fraction of their expected $v_\infty\sim 2800$km/s, and a comparatively low 
pre-shock wind velocity of $\sim 500$km/s follows~\citep{Rau05}.

WR stars are commonly thought to represent an evolved phase for the most massive stars in our Galaxy, however
the most luminous hydrogen-rich WR stars may still be thought of as the most massive, main-sequence stars~\citep{Mof06}, 
exhibiting strong, broad WR-like emission lines at optical wavelengths. Some WR stars are already established 
as non-thermal radio emitters, whereas claimed associations with $\gamma$-ray emission remain to be confirmed. 
It has been suggested that non-thermal radio emission observed in a significant fraction of the massive stars could 
be related to colliding winds in binaries~\citep{Dou00}. An important question is 
therefore whether all massive stars which produce non-thermal radio emission are indeed binaries. Another relevant 
issue is whether the relativistic electrons which produce the observed non-thermal (synchrotron) radio emission also 
produce detectable (inverse Compton) high-energy emission. Synchrotron emission has not been detected from the WR~20a system, 
presumably because of free-free-absorption in the optically thick stellar winds along the line of sight. 
However, WR~20a has been detected in X-rays~\citep{Mer96}. With a flux of the order of $10^{-12}$ erg cm$^{-2}$ s$^{-1}$, 
the soft X-ray emission is consistent with the predicted intensity for the thermal radiation from the 
shock-heated material in the stellar wind~\citep{Pit02}. However, non-thermal and thermal components of the X-ray emission 
are currently indistinguishable. In high-energy $\gamma$-rays, three unidentified EGRET sources have been found in the 
wider vicinity of RCW~49, 3EG~J1013-5915, 3EG~J1014-5705, and 3EG~J1027-5817~\citep{Har99}. These EGRET sources 
are characterized by both source confusion, and indication of extended emission, therefore viable counterparts are 
not established readily.  

It has been suggested that in massive binary systems high-energy $\gamma$-rays are produced either 
by optically-thin inverse Compton scattering of relativistic electrons with the dense photospheric 
stellar radiation fields in the wind-wind collision zone~\citep{Eic93, Whi95, Ben03, Rei06, Pit06}, 
or in hadronic emission scenarios. Here, $\gamma$-rays are either directly produced in neutral pion decays, 
with the mesons produced by inelastic interactions of relativistic nucleons with the wind 
material~\citep{Whi92, Ben03, Tor04, Rei06, Pit06}, or alternatively are produced in cascade models~\citep{Bed05}. 
In the latter scenario, $\gamma$-radiation originates from inverse-Compton pair cascades, which are 
initiated by high-energy neutral pion decay photons (from nucleon-nucleon interactions in the stellar winds). 
Due to the dense radiation field at the wind collision zone in the close binary WR~20a,  
$\gamma$-rays of a few tens of GeV are optically thick to photon-photon absorption, and pair 
cascades are unavoidable. If those nuclei reach sufficiently high energies (despite the expected weak shock 
in the extremely dense radiative environment), the projectile nucleons themselves are produced by photodisintegration 
of heavy nuclei in the radiation fields of the massive WR-stars~\citep{Bed05}. For a high magnetic field of order 
$10^3$G, particles could be energized up to $\sim 10^{15}$eV (adopting typical WR-wind parameters)
in the magnetic field reconnection scenario, and up to $10^3$GeV if first-order Fermi acceleration takes place there. 
The maximum power extractable from WR~20a must be a fraction of the kinetic wind energy 
provided by the system, estimated to be $\sim 10^{37}$erg/s.

Detectable VHE $\gamma$-radiation from the WR~20a binary system is so far only predicted in the pair cascade model, 
although detailed modeling of the WR~20a system in the other scenarios is still pending.  For VHE $\gamma$-rays, 
photon-photon absorption will diminish the observable flux from a close binary system such as WR~20a~\citep{Bed05,Dub06}.
  
Here we report on the first VHE $\gamma$-ray observations towards the WR binary WR~20a, and its immediate surroundings
with the H.E.S.S.\ telescopes. The observations and analysis yield the detection of a new source, HESS~J1023--575. 
This source is found to coincide with the central region of RCW~49, and accordingly, with the stellar cluster Westerlund~2, 
which hosts the WR binary WR~20a. The results and possible interpretation of these findings are presented in the following.

%
%

\section{H.E.S.S. Observations and Data Analysis}

The H.E.S.S. (High Energy Stereoscopic System) collaboration operates a telescope system of four 
imaging atmospheric Cherenkov telescopes located in Namibia (23$^\circ$16'17'' S 16$^\circ$29'58'' E), 
at 1800\,m above sea level. Each telescope has a tesselated spherical mirror with 13\,m diameter and 
$107$\,m$^2$ area, and a telescope spacing of 120 m. Cherenkov light emitted by extended air showers in 
the atmosphere is imaged by a high resolution camera (960 photo-multipliers, pixel size 0.16$^\circ$). 
Shower parameters and the primary particle type are determined by image reconstruction \citep{HESSCalib, HESSCrab}. 
The telescopes are operated in coincidence mode, which requires a trigger from at least two out of the 
four telescopes~\citep{HESSTrigger}. The H.E.S.S.\ telescope array achieves a point source sensitivity 
above 1\,TeV of $<2.0\times\,10^{-13}$\,cm$^{-2}$\,s$^{-1}$ (1\% of the flux from the Crab nebula) 
for a 5$\sigma$ detection in a 25\,hour observation.

The dataset described here consists of 14\,h (12.9\,h live time) of data taken between March and July 
2006, either on the nominal source location of WR~20a or overlapping data from the ongoing
Galactic plane survey. Quality selections were imposed on the data, excluding those taken during bad 
weather or with hardware irregularities. The data have been obtained in wobble-mode observations, 
where the telescopes are pointed offset from the nominal source location to allow for simultaneous 
background estimation. The wobble offsets for these observations range from 0.5$^\circ$ to 2$^\circ$, 
with the majority of data taken with wobble offset less than 0.8$^\circ$. The zenith angles range between 
36$^\circ$ and 53$^\circ$, resulting in an energy threshold of 380\,GeV for the analysis.

%
\begin{figure}[ht]
   \centering
   \includegraphics[width=0.48\textwidth]{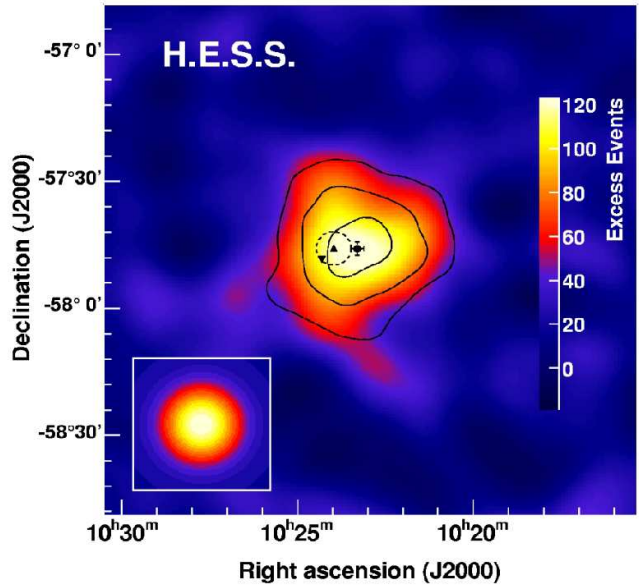} 
   \caption{Correlated excess sky map with an oversampling radius of 0.12$^\circ$ of the Westerlund~2/RCW~49 region. 
            The background in each bin is estimated using a ring around the test position. The map has been smoothed 
			with a two-dimensional Gaussian of radius 0.04$^\circ$ to reduce the effect of statistical fluctuations. 
			Overlaid contours correspond to statistical significances of 5, 7, and 9$\sigma$, respectively, determined 
			using the nominal oversampling radius for extended sources of 0.2$^\circ$. 
			The inlay in the lower left corner shows the excess distribution for a point-source derived from Monte Carlo 
			data with the same zenith-angle and offset distribution as the data. The cross denotes the best fit 
			position of the gamma-ray data with 1$\sigma$ statistical uncertainties. The WR binary WR~20a is indicated by 
			an upright filled triangle in the Westerlund~2 stellar cluster (dashed circle), the reversed filled triangle 
			denotes the location of WR~20b.}
   \label{Fig.1}
\end{figure}

%
\begin{figure}[ht]
   \centering
   \includegraphics[width=0.51\textwidth]{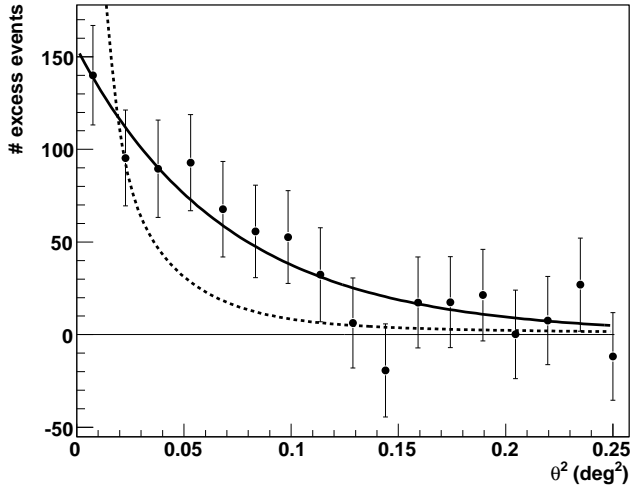} 
   \caption{Number of excess events versus the squared angular distance from the best fit position of the excess ($\theta^2$). 
            The dashed line shows the expectation for a point source derived from Monte Carlo data. The solid line is a 
			fit of the PSF folded with a Gaussian ($\sigma = 0.18^\circ \pm 0.02^\circ$).}
   \label{Fig.2}
\end{figure}

%
\begin{figure}[ht]
   \centering
   \includegraphics[width=0.48\textwidth]{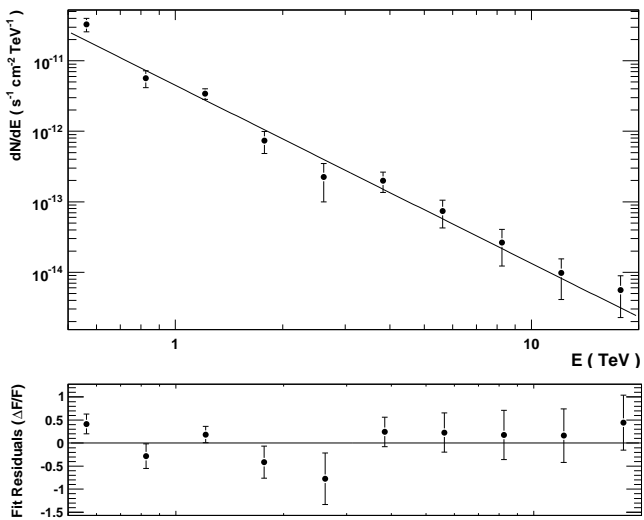} 
   \caption{Differential energy spectrum and fit residuals to a single power-law fit of HESS~J1023--575 from photons 
            inside the 85\% containment radius (0.39$^\circ$) around the best fit position. The background is estimated with 
			background regions of the same size and distance from the camera center as the signal region. The spectrum can be described 
			by a power law with a photon index of $\Gamma=2.53 \pm 0.16_{\mathrm{stat}} \pm 0.1_{\mathrm{syst}}$ and a normalization 
			at 1 TeV $\Phi_0 = (4.50 \pm 0.56_{\mathrm{stat}} \pm 0.90_{\mathrm{syst}}) \times 10^{-12}$\,TeV$^{-1}$\,cm$^{-2}$\,s$^{-1}$. 
			The integral flux for the whole excess is F(E$>$380\,GeV) = (1.3 $\pm$ 0.3) $\times 10^{-11}$\,cm$^{-2}$\,s$^{-1}$.}
   \label{Fig.3}
\end{figure}

The data have been analyzed using the H.E.S.S. standard Hillas analysis with standard cuts (image amplitude $>$ 80\,p.e.). 
Optical efficiency corrections have been applied as described in~\citet{HESSCrab}. A point source analysis on the nominal 
position of WR~20a resulted in a clear signal with a significance of 6.8 standard deviations. Further investigations of the 
skymap of this region revealed an extended excess with a peak significance exceeding 9 standard deviations (Fig.~1). The position 
of the center of the excess was derived by fitting the two-dimensional point spread function (PSF) of the instrument folded 
with a Gaussian to the uncorrelated excess map: $\alpha_{2000}$ = $10^{\rm h}23^{\rm m}18^{\rm s} \pm 12^{\rm s}$, 
$\delta_{2000}$ = -57$^\circ$45'50'' $\pm$ 1'30''. 
The systematic error in the source location is 20'' in both coordinates. The number of excess events versus the squared angular 
distance to this position in comparison to the expectation for a point source (dashed line) is shown in Fig.~2. The excess is clearly 
extended beyond the nominal extension of the PSF. A fit of a Gaussian folded with the PSF of the H.E.S.S. instruments gives an 
extension of $\sigma = 0.18^\circ \pm 0.02^\circ$. 
The differential energy spectrum for photons inside the corresponding 85\% containment radius of 0.39$^\circ$ is shown in Fig.~3. 
It can be described by a power law (dN/dE$= \Phi_0 \cdot (\mbox{E}/1\,\mbox{TeV})^{- \Gamma}$) 
with a photon index of $\Gamma=2.53 \pm 0.16_{\mathrm{stat}} \pm 0.1_{\mathrm{syst}}$ and a normalization at 1\,TeV of 
$\Phi_0 = (4.50 \pm 0.56_{\mathrm{stat}} \pm 0.90_{\mathrm{syst}}) \times 10^{-12}$\,TeV$^{-1}$\,cm$^{-2}$\,s$^{-1}$. 
The integral flux for the whole excess above the energy threshold of 380 GeV is (1.3 $\pm$ 0.3) $\times 10^{-11}$\,cm$^{-2}$\,s$^{-1}$. 
No significant flux variability could be detected in the data set. The fit of a constant function to the lightcurve binned 
in data segments of 28\,minutes has a chance probability of 0.14. The results were checked with independent analyses and 
were found to be in good agreement.

%
%

\section{HESS~J1023--575 in the context of $\gamma$-ray emission scenarios} 

The detection of VHE photons from the region RCW~49/Westerlund~2, characterized by a 
moderately hard power-law spectrum with an index $\Gamma$ $\sim$2.5 is indicative of 
the existence of extreme high-energy particle acceleration processes in this young 
($\sim$2-3 Myrs;~\cite{Pia98}) star forming region. The estimated luminosity 
above 380 GeV is $\sim 1.5\times 10^{35}$erg/s at a distance of 8~kpc, which 
corresponds to 0.5\% of the total kinetic energy available from the colliding winds of WR~20a, 
and $\sim$1.6\% of the kinetic energy of WR~20b. With a projected angular size of 
submilliarcsecond scale, the WR~20a binary system, including its colliding wind zone, would 
appear as a point source for observations with the H.E.S.S. telescope array. Unless there are 
extreme differences in the spatial extent of the particle distributions producing 
radio, X-ray, and VHE $\gamma$-ray emission, \emph{scenarios based on the colliding stellar winds
in the WR~20a binary system} face the severe problem of accounting for a source extension of 
0.18$^\circ$ in the VHE waveband. At a nominal distance of WR~20a (8.0kpc;~Rauw et al. 2007), this source 
extension is equivalent to a diameter of 28 pc for the emission region, consistent in size with 
theoretical predictions of bubbles blown from massive stars into the ISM~\citep{Cas75}. The spatial extension
found for HESS~J1023--575 contradicts emission scenarios where the bulk of the $\gamma$-rays are produced 
close to the massive stars. If VHE $\gamma$-rays were indeed produced near the massive stars in the binary, 
a regular modulation of the $\gamma$-ray flux due to absorption would be expected~\cite{Dub06}. Future observations 
of HESS~J1023--575 will allow one to probe for such a pattern, which could discard or unambiguously label the origin 
of the VHE emission.
 
Alternatively, the emission could arise from \emph{collective effects of stellar winds in the Westerlund~2 cluster}. 
Diffusive shock acceleration in cases where energetic particles experience multiple shocks~\citep{Kle00} can be considered
for Westerlund~2. The stellar winds may provide a sufficiently dense target for high-energy particles (accelerated to 
hundreds of TeV in the winds themselves or driven by supernova explosions), allowing the production of 
$\pi^0$-decay $\gamma$-rays via inelastic pp-interactions. According to \citet{Rau07} Westerlund~2 may host about 
4500 M$_{\sun}$ in the form of stars with M $>$ M$_{\sun}$. The mechanical energy injected through the stellar winds determines 
mainly the total energy available for collective wind effects, estimated to be about $5.4\times 10^{37}$erg/s at most. 
A collective wind scenario proposed by~\citet{Dom06} predicts that the extension 
of a $\gamma$-ray source corresponds to the volume filled by the hot, shocked stellar winds. Variability is generally not 
expected since flux modulation of the winds of individual stars or binaries is believed to average out in a collective wind.  
Interestingly, a putative VHE source may not possess a non-thermal counterpart at MeV/GeV energies in such a scenario, 
since convection will prevent low-energy particles from entering the wind. In both colliding wind zone and collective 
wind models, one would expect sub-TeV spectra that resemble those of Supernova remnants. 

\emph{Magneto-hydrodynamic (MHD) particle acceleration} (e.g. by multiple shocks or turbulence 
produced by supersonic flows) in a magnetized plasma may also be considered when particles penetrate 
into a dense medium. In such a case, the massive stellar winds of Westerlund~2 could accomplish three tasks: 
they could ensure sufficient particle injection into the turbulent plasma, feed the magnetic turbulence with 
energy via wind-wind interactions in the massive star association, or provide copious photons and dense material 
to serve as target for TeV-photon producing particle-photon and photon-photon interactions. 
\begin{figure}[ht]
   \centering
   \includegraphics[width=0.48\textwidth]{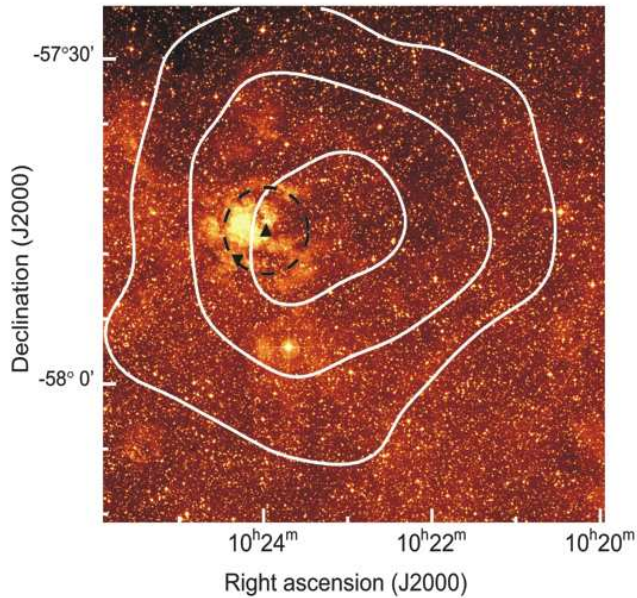} 
   \caption{HESS~J1023--575 significance contours (corresponding 5, 7 and 9 $\sigma$), overlaid 
   on a B-band image from the Second Palomar Observatory Sky Survey (POSS-2). The filled circle 
   denotes the best fit position with 1$\sigma$ statistical uncertainties. The WR binary WR~20a is indicated 
   by an upright filled triangle in the Westerlund~2 stellar cluster (dashed circle), the reversed filled 
   triangle denotes the location of WR~20b.}
   \label{Fig.4}
\end{figure} 
Supershells, molecular clouds, and inhomogeneities embedded in the dense hot medium may serve as the targets for $\gamma$-ray production in Cosmic Ray 
interactions. Such environments have been already studied in the nonlinear theory of particle acceleration by large--scale 
MHD turbulence \citep{Byk87}. Shocks and MHD turbulent motion inside a stellar bubble or superbubble can efficiently transfer 
energy to cosmic rays if the particle acceleration time inside the hot bubble is much shorter than the bubble's expansion time. 
X-ray observations of RCW 49 revealed, apart from point-like emission from individual stars, 
a diffuse component centered on the ionizing cluster, widely extending outside the core, with 
a soft component and a hard tail~\citep{Bel94, Tow05}. This indicates the existence of a wind-blown 
shell from the stars at the cluster core, filled with a hot ($\sim 0.1-3$ keV) tenuous plasma 
shocked by turbulence. Two wind-blown bubbles are revealed from continuum radio observations: one around WR~20b with a 
diameter of $\sim 4.1\arcmin$ and an expansion velocity of 67~km/s~~\citep{Sha91}, the other around the Westerlund~2 
cluster center with a diameter of $\sim 7.3\arcmin$~\citep{Whi97}. A ridge of enhanced GHz radio emission lies between 
these two shells, and it has been suggested that this may represent a site of collision between these two shells~\citep{Whi97}. 
Most interestingly, the extent of a blister~\citep{Whi97} on the western side of the Westerlund~2 bubble appears to be compatible 
in direction and location with the center of gravity of HESS~J1023--575. Such a blister is indicative for rapid expansion 
into a low-density medium outside the wind-blown bubble.
\begin{figure}[ht]
   \centering
   \includegraphics[width=0.48\textwidth]{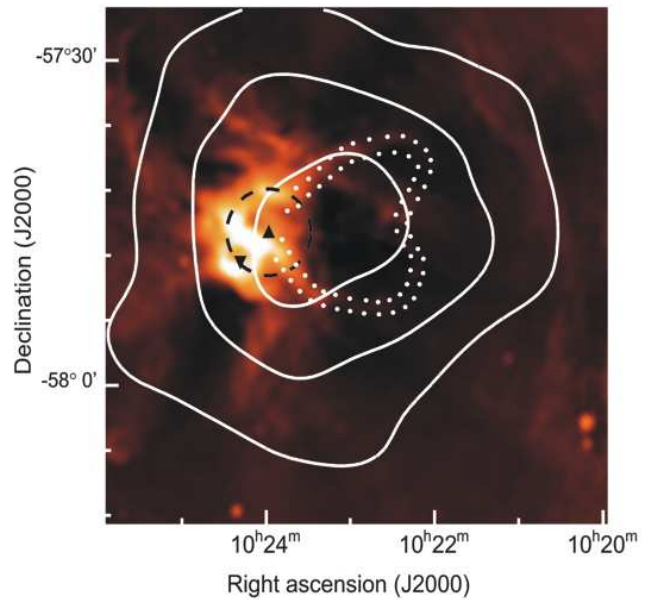} 
   \caption{HESS~J1023--575 significance contours (corresponding 5, 7 and 9 $\sigma$), overlaid 
   on a 843 MHz image from the Molonglo Observatory Synthesis Telescope (MOST). Black symbols and
   dashed circle as in Fig.~1. The wind-blown bubble around WR~20a, and the blister to the west of 
   it are seen as depressions in the radio continuum map. The blister is indicated by white dots as in \citet{Whi97},
   and appears to be compatible in direction and location with the center of gravity of HESS~J1023--575.}
   \label{Fig.5}
\end{figure} 
Figs.~4 and 5 show overlays of the detection significance contours of HESS~J1023--575 (E$>$380 GeV) 
on the B-band image of the region as obtained from the Second Palomar Observatory Sky Survey (POSS-2), and
on the 843 MHz radio continuum image from the Molonglo Observatory Synthesis Telescope (MOST), respectively.  
Whereas the open cluster Westerlund~2 does not perfectly coincide with the peak of the emission seen at 
very high energy $\gamma$-rays (Fig.~4), the direction of the outbreak and spatial location of the blister 
is a better match (Fig.~5). The later association also relaxes constraints arising of the detection of 
HESS~J1023--575 as an extended VHE $\gamma$-ray emitter.
 
\emph{Shock acceleration at the boundaries of the blister} may enable particles to diffusively re-enter into 
the dense medium, thereby interacting in hadronic collisions and producing $\gamma$-rays. Consequently, 
a scenario as outlined in \citet{Vol83} for a supernova-driven expansion of particles into a low density medium 
may be applicable to the expanding stellar winds into the ambient medium. If one accepts such a scenario here, 
it might give the first observational support of $\gamma$-ray emission due to diffusive shock acceleration from 
supersonic winds in a wind-blown bubble around WR~20a, or the ensemble of hot and massive OB stars from a 
superbubble in Westerlund~2, breaking out beyond the edge of a molecular cloud~\citep{Ten79, Vol82, Ces83, Byk01}. 
Accordingly, one has to consider that such acceleration sites will also contribute to the observed flux of 
cosmic rays in our Galaxy~\citep{Cas80}.

Other open star clusters hosting WR-stars are known to supersede the total injected mechanical energy 
from winds of massive stars as available in Westerlund~2: Westerlund~1, which is older than Westerlund~2 and 
lies at a recently revised distance of 5~kpc \citep{Cro06}, and NGC~3603, a very prominent galactic starburst 
region at a distance of 7~kpc. Both offer approximately $10^{39}$erg/s \citep{Mun06, Ste03}. 
Search for TeV-emission from open stellar clusters has been previously carried out for a list of northern clusters 
by HEGRA~\citep{HEGRA}, resulting in upper limits in the order of $10^{-11}$ to $10^{-13}$\,cm$^{-2}$\,s$^{-1}$ at 
varying energy thresholds between 0.8 and 3 TeV. The southern hemisphere stellar cluster Westerlund~2 was not targeted 
by HEGRA. The detection of very-high-energy $\gamma$-rays associated with Westerlund~2 by H.E.S.S. clearly motivates 
systematic searches for TeV-emission from massive open stellar clusters.

%
%

\section{Conclusions}

H.E.S.S. observations have led to the discovery of an extended source HESS~J1023--575 in the direction of 
Westerlund~2. Possible sources are the massive WR binary system WR~20a (although source extension and variability 
studies do not support a colliding wind scenario at present), the young stellar cluster Westerlund~2 (although the 
cluster itself appears too compact to account for the observed VHE emission in a collective wind scenario), and 
cosmic rays accelerated in bubbles or at their termination shock and interacting with their environment. 
Further observations with the H.E.S.S. telescope array will help to discriminate among the alternatives in the 
interpretation of the observed VHE $\gamma$-ray emission.

However, the convincing association of HESS~J1023--575 with a new type of astronomical object---a massive HII region 
and its ionizing young stellar cluster---profoundly distinguishes this new detection by the H.E.S.S. telescope array 
already from other source findings made during earlier Galactic Plane Scan observations~\citep{HESSScanI, HESSScanII}.

\begin{acknowledgements}
  The support of the Namibian authorities and of the University of
  Namibia in facilitating the construction and operation of H.E.S.S.\
  is gratefully acknowledged, as is the support by the German Ministry
  for Education and Research (BMBF), the Max Planck Society, the
  French Ministry for Research, the CNRS-IN2P3 and the Astroparticle
  Interdisciplinary Programme of the CNRS, the U.K. Particle Physics
  and Astronomy Research Council (PPARC), the IPNP of the Charles
  University, the South African Department of Science and Technology
  and National Research Foundation, and by the University of
  Namibia. We appreciate the excellent work of the technical support
  staff in Berlin, Durham, Hamburg, Heidelberg, Palaiseau, Paris,
  Saclay, and in Namibia in the construction and operation of the
  equipment.\\
  Usage of the Digitized Sky Survey (DSS-2) at ESO/ST-ECF, and the Molonglo 
  Galactic Plane Survey (MGPS) at the IoA, University of Sydney is acknowledged.\\
  The authors wish to thank the referee, Anthony Moffat, for his very constructive 
  and supportive feedback during refereeing.\\
\end{acknowledgements}

\bibliographystyle{aa}

\end{document}